\pdfoutput=1

\documentclass[11pt]{article}

\usepackage[preprint]{coling}

\usepackage{times}
\usepackage{latexsym}
\usepackage{booktabs}
\usepackage{amsmath}
\usepackage{appendix}
\usepackage[T1]{fontenc}

\usepackage[utf8]{inputenc}

\usepackage{microtype}

\usepackage{inconsolata}

\usepackage{graphicx}

\title{LLM Confidence Evaluation Measures in Zero-Shot CSS Classification}

\author{
  \textbf{David Farr \textsuperscript{1,2}},
  \textbf{Iain Cruickshank \textsuperscript{3}},
  \textbf{Nico Manzonelli \textsuperscript{2}},
  \textbf{Nicholas Clark \textsuperscript{1}},
  \textbf{Kate Starbird \textsuperscript{1}},
  \textbf{Jevin West\textsuperscript{1}}
\\
\\
\\
  \textsuperscript{1}University of Washington,
  \textsuperscript{2}Army Cyber Technology and Innovation Center,
  \textsuperscript{3}Carnegie Mellon University
\\
  \small{
    \textbf{Correspondence:} \href{mailto:dtfarr@uw.edu}{dtfarr@uw.edu}
  }
}

\begin{document}
\maketitle
\begin{abstract}
Assessing classification confidence is critical for leveraging large language models (LLMs) in automated labeling tasks, especially in the sensitive domains presented by Computational Social Science (CSS) tasks. In this paper, we make three key contributions: (1) we propose an uncertainty quantification (UQ) performance measure tailored for data annotation tasks, (2) we compare, for the first time, five different UQ strategies across three distinct LLMs and CSS data annotation tasks, (3) we introduce a novel UQ aggregation strategy that effectively identifies low-confidence LLM annotations and disproportionately uncovers data incorrectly labeled by the LLMs. Our results demonstrate that our proposed UQ aggregation strategy improves upon existing methods and can be used to significantly improve human-in-the-loop data annotation processes.
\end{abstract}

\section{Introduction}

Large Language Models (LLMs) have transformed the way artificial intelligence is integrated into professional workflows, with applications that span healthcare \cite{health}, academia \cite{academics}, cybersecurity \cite{cyber}, software development \cite{software}, and many others. However, research shows that users struggle to identify incorrect LLM responses which poses a problem because LLMs are less likely to refrain from answering questions they do not know as they scale with size and complexity \cite{zhou2024larger}. Despite these challenges, LLMs have proven effective in synthesizing vast amounts of data and applying contextual understanding, making them a popular choice for integration into natural language processing tasks, particularly in zero-shot classification settings where prior training data is unavailable \cite{zeroshot}. 

With broad applications in critical industries, LLM-generated responses that are assumed to be correct can lead to drastic second- and third-order consequences when answered incorrectly and integrated into decision-making processes. Although some LLMs incorporate expressions of uncertainty \cite{tian2023just}, developers often restrict the output of the model to a predetermined set of responses to manage nondeterministic behavior or reduce token generation cost \cite{constrained}. However, these constraints can cause LLMs to provide confident answers even when they lack the correct knowledge. While LLMs are useful for large-scale data annotation tasks, there remains uncertainty as to which labels are correct or how to best quantify label confidence in LLM-generated annotations, especially in multi-modal systems.

This paper evaluates various Uncertainty Quantification (UQ) methods to assess LLM confidence in data annotation tasks applied to Computational Social Science (CSS) problems. Based on these results, we present a simple UQ aggregation strategy to help identify misclassified LLM-labeled data. We constrain our settings to realistic industry scenarios where previously labeled data is unavailable to simulate common, real-world problems. Additionally, we propose a new evaluation metric that assesses the recall of misclassified LLM-labeled data at low-confidences and compare UQ techniques using the Area Under Curve (AUC) analysis by applying thresholds based on percentiles of confidence scores. Our methodology has significant implications for systems that use human-machine teaming for data annotation tasks by better identifying data on which humans should spend finite resources.

\section{Related Works}

\citealp{zhou2024larger} show that as LLMs scale, they become more confident and less avoidant in answering questions. However, this increased confidence comes at a cost: they answer questions incorrectly more frequently compared to smaller LLMs, which were more likely to avoid answering altogether. In a related study, \citealp{constrained} demonstrate the importance of constraining LLM outputs in software development workflows to ensure predictability. Together, these works highlight both the internal challenge of larger LLMs being more prone to incorrect answers instead of avoidance, as well as the common practice of imposing constraints on LLM outputs to improve workflow predictability.

In the field of LLM UQ techniques, \citealp{supervised1} demonstrates an effective method of UQ via supervised calibration from utilizing hidden activation layers. \citealp{humancollab} integrate a human annotated training set to train an external BERT-based verifier to select data that the LLM was likely to mislabel for later external human annotation. However, these methods require a labeled dataset for training an external supervised ML model which is not available in many contexts.
 
As such, recent research has investigated zero-shot UQ techniques for LLMs. \citealp{kadavath2022languagemodelsmostlyknow} and \citealp{tian2023just} show that an effective technique to assess confidence in LLMs tuned with reinforcement learning human feedback (RLHF) is prompting the model to evaluate its confidence in its own answer. \citealp{kumar2023conformalpredictionlargelanguage} find that the uncertainty estimates from conformal prediction are closely correlated with the accuracy of the prediction. 

Instead of relying on the LLM to self-report confidence, other approaches analyze model output. For example, \citealp{entropy} show that the approximation of entropy using measurements on a restricted set of returned tokens is a valid mechanism to assess confidence in multiple-choice questions. Additionally, \citealp{redct} present an effective mechanism for identifying mislabeled data is using the absolute difference between the two highest log probability values returned. Finally, \citealp{semanticuq} look for semantic differences in responses can inform uncertainty. Our work builds on the existing literature by comparing a sample of the aforementioned UQ mechanisms for zero-shot classification while proposing a new methodology that takes advantage of existing UQ techniques through an ensemble method.

\section{Uncertainty Quantification Techniques}

In this section we describe the five UQ techniques used in the study. 

\subsection{Quantitative and Qualitative Self-report }

Our first and second UQ techniques are driven by the work of \citealp{tian2023just}, who show that RLHF tuned LLMs can self-assess answer confidence. We accomplish this by prompting the model to give a quantitative assessment of its confidence on a scale between 0 and 100. We also assess the ability of language models to map its uncertainty in qualitative terms. Our hypothesis being that open-source LLMs may perform better using normal language as opposed to probabilistic quantitative values. We accomplish this by asking models to report either \textit{no}, \textit{low}, \textit{medium}, \textit{high}, or \textit{absolute} confidence in their responses. Then we map those responses to quantitative values of 0, 0.25, 0.50, 0.75, and 1 to allow comparability to other confidence measures. Access to all prompt examples and datasets can be found in the availability section.

\subsection{Confidence Score}

We use the confidence score method from \citealp{redct}, where the authors define the confidence score as the absolute value of the difference between the highest token label log probability and the second-highest token label log probability within a constrained set of tokens. Let \(\mathcal{T}\) represent the set of given tokens, and \(P(t)\) denote the distribution of log probabilities across each token \(t \in \mathcal{T}\). The log probability is then computed using the formula

\begin{equation}
    \text{\(C\)} = \left| \max_{t \in \mathcal{T}} P(t) - \max_{t \in \mathcal{T} \setminus \{t^*\}} P(t) \right|,
\label{csequation}
\end{equation}

\noindent where \(t^*\) is the token corresponding to the highest probability \( \max_{t \in \mathcal{T}} P(t) \). We refer to this metric as \textit{C\_score} in our results section.

\subsection{Log Inverse}

We additionally test a commonly used method to convert the logarithmic probability of the highest returned token into a probability. This allows us to investigate whether the difference between the confidence score (based on the top two token probabilities) and the direct probability of the highest token leads to significant differences in sampling outcomes. Specifically, let \( t^* \) represent the token with the highest probability, and let \( \log P(t^*) \) denote the log probability of this token. The probability for the token \( t^* \) is obtained by exponentiating the log probability.

For our results, we refer to this methodology as the \textit{log inverse}.

\subsection{Confidence Ensemble}

Finally, we introduce the following UQ aggregation strategy, which is more resource intensive than the previous three, requiring the aforementioned confidence score from multiple LLMs, but is meant to reward LLMs for converging on a single label, while not penalizing a divergence of LLM-responses. This is especially important in classification tasks with multiple target classes.

Let \( \mathcal{L} \) represent the set of LLMs, and for each LLM \( L_i \), the token with the highest probability is denoted by \( t_{L_i}^* \), and the corresponding confidence score \( C_{L_i} \) is given in Equation \ref{csequation}. To aggregate confidence scores when multiple LLMs provide the same answer \( t^* \), the overall confidence score \( C_{\text{agg}} \) is calculated as

\begin{equation}
    C_{\text{agg}} = \sum_{\{L_i \in \mathcal{L} \,|\, t_{L_i}^* = t^*\}} C_{L_i},
\label{csagg}
\end{equation}

\noindent where \( t^* \) is the common token predicted by the LLMs. For our results section, this methodology is referred to as \textit{C\_ensemble}.

\section{Experimental Design}

We evaluated our five UQ techniques across three different LLMs and three distinct CSS tasks. For each LLM and task, we rank all annotated data from least confident to most confident, allowing us to sample low-confidence data for human-in-the-loop labeling or high-confidence data for downstream classifiers. Each CSS task is pulled from common benchmark datasets for stance, ideology, and frame detection. For stance detection, we use the SemEval-2016 dataset \cite{StanceSemEval2016}. For ideology detection, we use the ideological books corpus (IBC) from \cite{sim-etal-2013-measuring} with sub-sentential annotations \cite{iyyer2014neural}. For frame detection, we use the Gun Violence Frames Corpus (GVFC) from \cite{gvfc}. The LLMs chosen were Llama-3.1 8B Instruct, Flan UL2, and GPT-4o. This selection was intentional to show a variety of parameter sizes and the integration of a RLHF-tuned model to show utility in sampling strategy mechanisms across different LLMs.

\subsection{Evaluation Metric}

\begin{figure*}[h!]
    \centering
    \includegraphics[scale=.45]{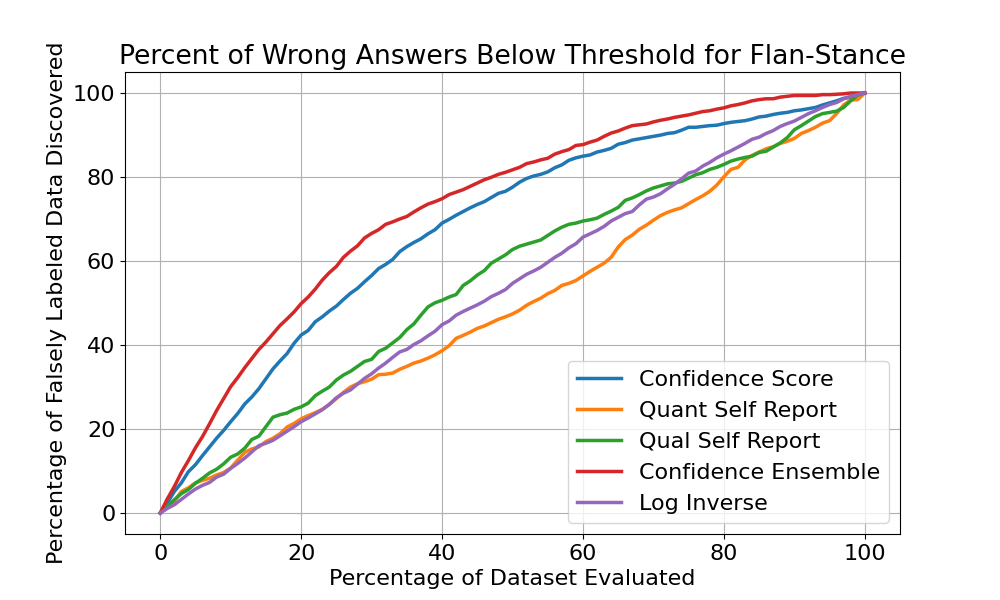}
    \caption{Graph depicts the percent of incorrect data annotations identified given the amount of data sampled for stance detection via Flan UL2. This shows we can find approximately half of all incorrect data annotations by checking only the bottom 20\% of data evaluated by our confidence ensemble method. This graph also is meant to show a natural understanding of why AUC is a valuable measure for uncertainty quantification when measuring by percent of false labels detected. }
    \label{flan_graph}
\end{figure*}

In order to demonstrate the efficacy of each UQ strategy, we devise a metric that measures a confidence scoring techniques' ability to recall misclassified LLM-labeled data at low-confidences. When used to inform sampling for human-in-the-loop labeling, we would like to send a small sample of LLM labeled data to human data annotators for evaluation. Ideally, human evaluation is only applied to the data that the LLM is likely to misclassify, which boosts overall classification accuracy under the assumption that the human will correctly label data misclassified by the LLM. By selecting data based on the lowest percentile of confidence scores, we aim to select misclassfied examples for humans to evaluate. Therefore, we measure the percentage of falsely LLM-labeled data recalled as a function of the percentage of the total dataset evaluated based on the same the bottom percentile of confidence scores. This curve is depicted in Figure \ref{flan_graph}. 

Our goal is to succinctly measure performance across UQ techniques, LLMs, and datasets. In order to accomplish this, we report the Area Under Curve (AUC) of the proportion of wrong examples to evaluated examples. As indicated in Figure \ref{flan_graph}, a higher AUC indicates a better UQ-informed, data sampling strategy. The AUC is calculated for the curve evaluating the dataset from none of the dataset to the full dataset. All graphs for evaluated models, sampling strategies, and datasets are shown in Appendix \ref{plots}. We also report the accuracy on each task for the LLM evaluated on the labeling tasks in Appendix \ref{labelresults}.

\begin{table*}[h!]
\begin{tabular}{|l|lll|lll|lll|l|}
\hline
 & \multicolumn{3}{c|}{GPT} & \multicolumn{3}{c|}{Flan} & \multicolumn{3}{c|}{Llama} &  \\ \hline
UQ Metric & \multicolumn{1}{c|}{Stance} & \multicolumn{1}{c|}{IBC} & \multicolumn{1}{c|}{GVFC} & \multicolumn{1}{c|}{Stance} & \multicolumn{1}{c|}{IBC} & \multicolumn{1}{c|}{GVFC} & \multicolumn{1}{c|}{Stance} & \multicolumn{1}{c|}{IBC} & \multicolumn{1}{c|}{GVFC} & \multicolumn{1}{c|}{\textbf{AVG}} \\ \hline
Qual. & \multicolumn{1}{l|}{64.6} & \multicolumn{1}{l|}{60.8} & 55.1 & \multicolumn{1}{l|}{55.9} & \multicolumn{1}{l|}{52.5} & 50.1 & \multicolumn{1}{l|}{56.5} & \multicolumn{1}{l|}{50.6} & 54.6 & 55.6 \\ \hline
Quant. & \multicolumn{1}{l|}{66.3} & \multicolumn{1}{l|}{64.6} & 59 & \multicolumn{1}{l|}{49.8} & \multicolumn{1}{l|}{50.6} & 46.25 & \multicolumn{1}{l|}{51.6} & \multicolumn{1}{l|}{51.7} & 55.3 & 55.0 \\ \hline
Log Inverse & \multicolumn{1}{l|}{57.4} & \multicolumn{1}{l|}{42.4} & 66.7 & \multicolumn{1}{l|}{53.5} & \multicolumn{1}{l|}{\textbf{60.8}} & 63.9 & \multicolumn{1}{l|}{60.3} & \multicolumn{1}{l|}{56.4} & 60.1 & 57.9 \\ \hline
C\_Score & \multicolumn{1}{l|}{67.1} & \multicolumn{1}{l|}{63.3} & \textbf{67.2} & \multicolumn{1}{l|}{68.1} & \multicolumn{1}{l|}{60.3} & 62.7 & \multicolumn{1}{l|}{66.5} & \multicolumn{1}{l|}{\textbf{62.7}} & 59.6 & 64.2 \\ \hline
C\_Ensemble & \multicolumn{1}{l|}{\textbf{71.4}} & \multicolumn{1}{l|}{\textbf{65.0}} & 66.6 & \multicolumn{1}{l|}{\textbf{73.2}} & \multicolumn{1}{l|}{54.3} & \textbf{69.3} & \multicolumn{1}{l|}{\textbf{73.6}} & \multicolumn{1}{l|}{58.4} & \textbf{69.1} & \textbf{66.8} \\ \hline
\end{tabular}
\caption{Depicts the Area Under Curve (AUC) metric across selected sampling strategy, dataset, and language model. The top performing sampling strategy for each task is in bold. We also report the average performance for each sampling strategy. Across all LLM types, the confidence ensemble method shows the most robustness.}
\label{results_table}
\end{table*}

\section{Results}

Our results are shown in Table \ref{results_table}. Overall, the confidence ensemble uncertainty quantification measure is the most robust evaluated UQ strategy, proving to be effective across all model types. In the RLHF model evaluated, GPT-4o, quantitative self-reporting seemed also to be an effective strategy. Interestingly, for GPT-4o the log inverse performance did not closely resemble the confidence score or ensemble metrics. In the evaluated data, GPT appeared to return less deterministic responses, meaning that it was not as likely to achieve a high log inverse score when searching for a selected token, even when the model found it to be an easy task when evaluated using our other UQ techniques. On the contrary, the difficulty or ease of the tasks is highlighted in non-deterministic models with deterministic constraints by looking for the distribution between constrained tokens. For our non-RLHF models, Flan and Llama, our results indicate that self-assessment is a poor strategy; however, if underlying token log probabilities are not available, they seem to perform better when asked to qualitatively assess their confidence as opposed to answering with a numeric response. Like GPT-4o, the confidence ensemble appears to be the most robust metric, followed by the confidence score.

\section{Conclusion}
Through this work, we have evaluated several easy-to-implement UQ-based sampling strategies for finding erroneously LLM-labeled data in a zero-shot setting (i.e., common data annotation setting). We find that using confidence ensembles is the most effective mechanism for discovering erroneously labeled data. When only one LLM is being implemented, using the underlying distribution between the top two log probabilities is also an effective UQ mechanism. Using LLMs to label CSS data is a rapidly growing trend; however, it is important for humans to assess the quality of the labels generated. Our UQ strategies show that we can find a disproportionate amount of incorrectly annotated data, which should be evaluated by humans, by looking at small quantities driven by uncertainty quantification. 

\section{Availability and Resources}
All code and data to produce these experiments can be found at https://anonymous.4open.science/r/UQMetrics-E69C. Two NVIDIA A6000 GPUs were used over the course of 18 hours for local LLMs. GPT was used to debug analysis graphs.

\section{Limitations}
We have only tested this methodology on three different datasets and three LLMs. Although it has seemingly extrapolated across the nine different testing combinations for the five separate sampling strategies, like all methodologies, it would benefit from testing across additional models and datasets for increased robustness. Furthermore, while we tested against different tasks, they were all broadly in the CSS space and against a constrained set of choices. For additional applications or labeling settings, more testing would need to be done. Finally, our most effective strategy required access to more than one LLM and underlying token log probability values, a combination that, while common, is not ubiquitous.

\bibliography{custom}

\appendix{}
\section{LLM Annotation Accuracy}\label{labelresults}
The table below shows the accuracy of LLM labeling for the three tasks given for each LLM evaluated.

\begin{table}[h!]
\begin{tabular}{|l|l|l|l|}
\hline
       & FLAN UL2 & GPT-4o & Mistral 8b \\ \hline
Stance & 75.6     & 77.4   & 72.4       \\ \hline
IBC    & 62.3     & 62.5   & 65.2       \\ \hline
GVFC   & 58.7     & 69.5   & 58.3       \\ \hline
\end{tabular}
\end{table}

\section{Plots}\label{plots}
Below are all plots associated with reported AUC metrics.

\begin{figure*}[h]
    \centering
    \includegraphics[scale=.50]{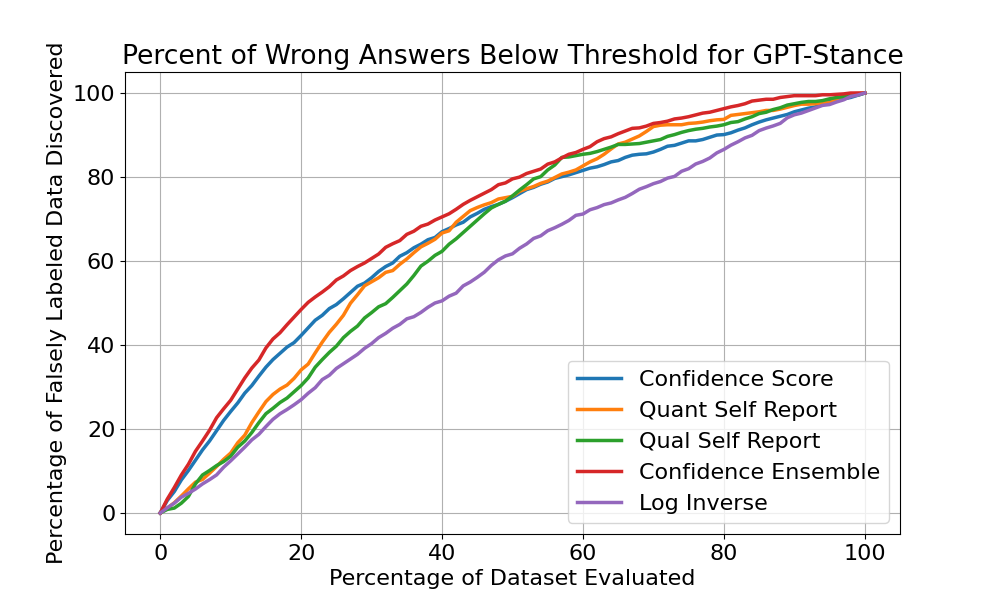}
\end{figure*}

\begin{figure*}[h]
    \centering
    \includegraphics[scale=.50]{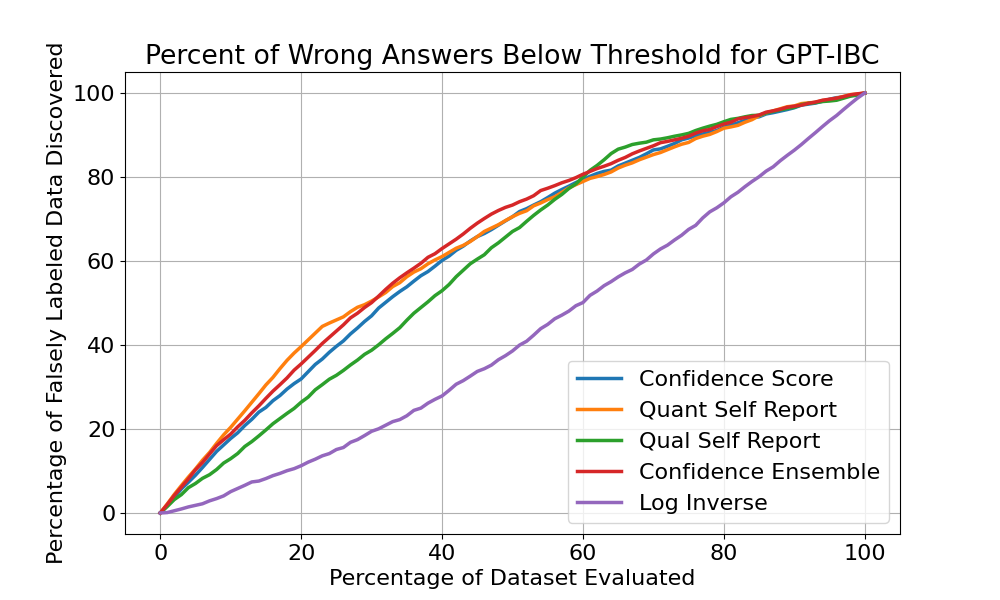}
    \includegraphics[scale=.50]{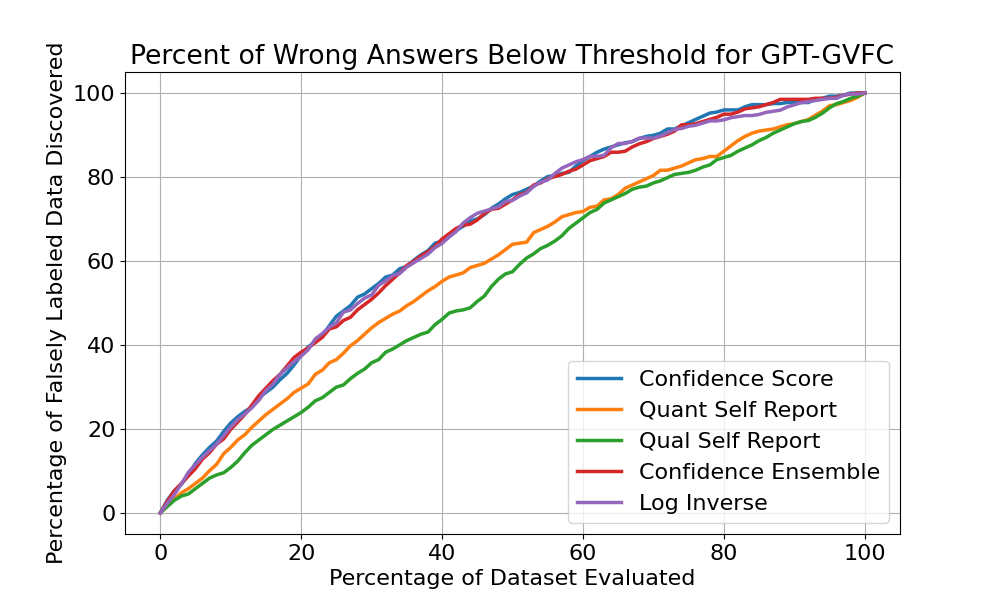}
\end{figure*}

\begin{figure*}[h]
    \centering
    \includegraphics[scale=.50]{flan-stance.png}
    \includegraphics[scale=.50]{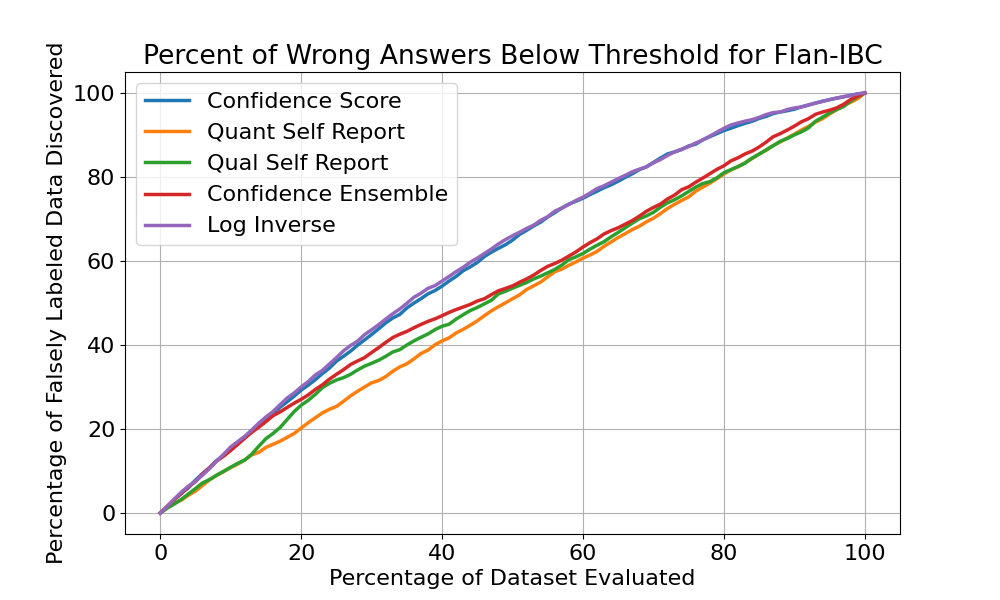}
    \includegraphics[scale=.50]{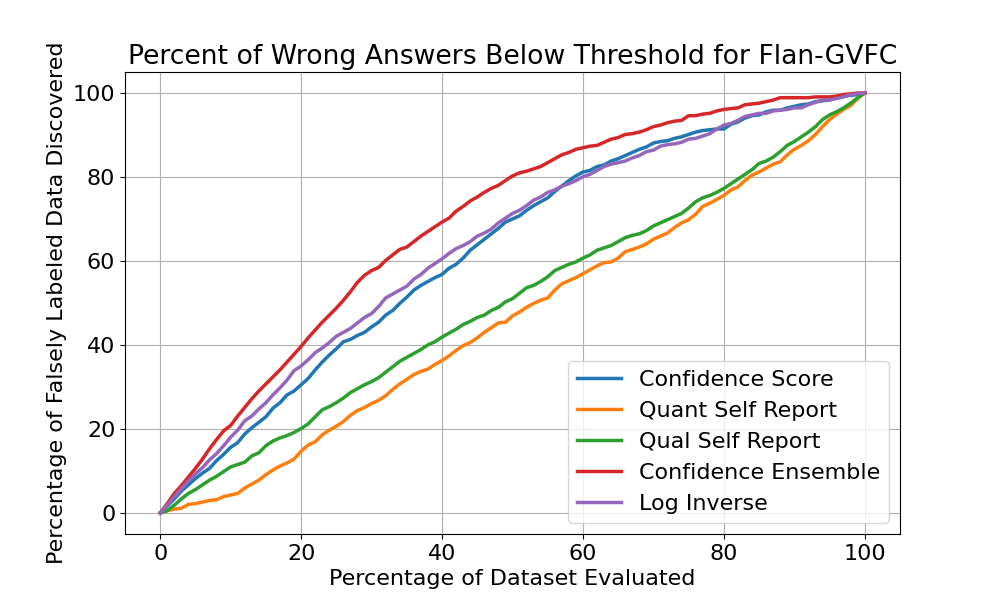}
\end{figure*}

\begin{figure*}[h]
    \centering
    \includegraphics[scale=.50]{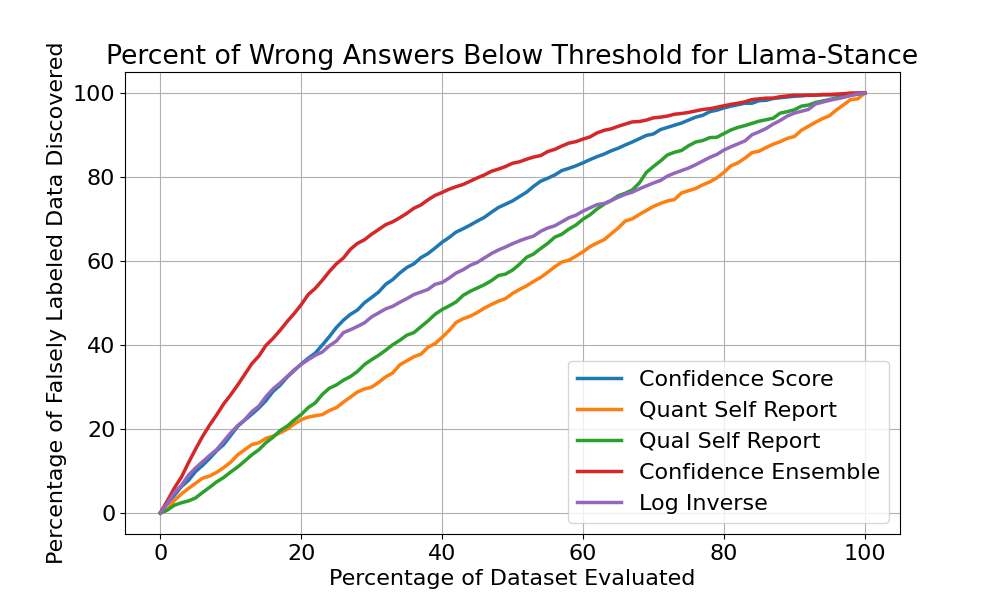}
    \includegraphics[scale=.50]{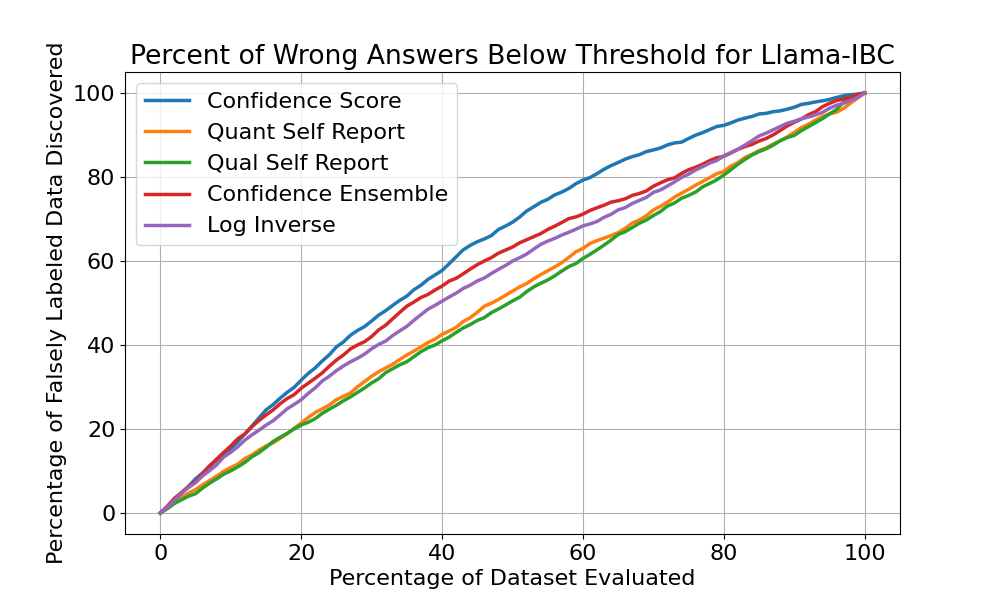}
    \includegraphics[scale=.50]{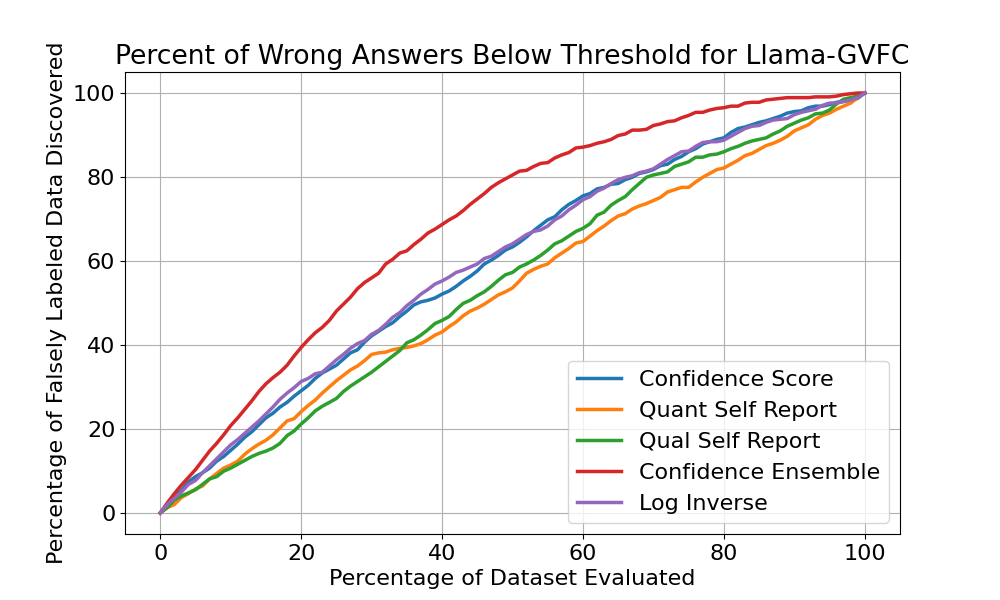}
\end{figure*}

\end{document}